\documentclass[12pt]{iopart}
\usepackage{epsf}  
\begin{document}

\article[Gapless phases of color-superconducting matter]
{Strangeness in Quark Matter 2004}
{Gapless phases of color-superconducting matter}

\author{Igor A. Shovkovy\dag\footnote[3]{Speaker at the 
conference. On leave from Bogolyubov 
Institute for Theoretical Physics, 03143, Kiev, Ukraine.},\
Stefan B.\ R\"uster\ddag\
\ and Dirk H.\ Rischke\ddag}

\address{\dag\ Frankfurt Institute for Advanced Studies, 
Johann Wolfgang Goethe-Universit\"{a}t, 
D-60054 Frankurt am Main, Germany}
\ead{shovkovy@th.physik.uni-frankfurt.de}\vspace{3mm}

\address{\ddag\ Institut f\"ur Theoretische Physik, 
Johann Wolfgang Goethe-Universit\"{a}t, 
D-60054 Frankurt am Main, Germany}

\date{10 November, 2004}

\begin{abstract}
We discuss gapless color superconductivity for neutral quark matter 
in $\beta$ equilibrium at zero as well as at nonzero temperature. 
Basic properties of gapless superconductors are reviewed. The 
current progress and the remaining problems in the understanding 
of the phase diagram of strange quark matter are discussed.

\end{abstract}

\pacs{12.38.-t, 12.38.Aw, 12.38.Mh, 26.60.+c}


\submitto{\JPG}


\section{Introduction}

The estimated central densities of compact stars could be 
sufficiently large to support the existence of deconfined 
quark matter. Such matter should develop a Cooper 
instability with respect to diquark condensation, and become 
color superconducting \cite{BailinLove,cs,cfl,weak}. Note that 
typical temperatures inside compact stars are so low that the 
diquark condensate, if produced, would not melt.

Matter in the bulk of a compact star should be neutral (at least, on 
average) with respect to electric as well as color charges. Otherwise, 
the star would not be bound by gravity which is much weaker than 
electromagnetism. Matter should also remain in $\beta$ equilibrium. 
The latter requires that the rate 
of the $\beta$-decay processes (i.e., $d \to u + e^{-} + \bar\nu_{e}$ 
and $s \to u + e^{-} + \bar\nu_{e}$) should be equal to the rate of 
the corresponding electron capture processes (i.e.,
$u + e^{-} \to d + \nu_{e}$ and $u + e^{-} \to s + \nu_{e}$).

After the charge neutrality and the $\beta$-equilibrium conditions 
are enforced, the chemical potentials of different quarks satisfy 
relations that may interfere with the dynamics of Cooper 
pairing. If this happens, some color-superconducting phases may 
become less favored than others. For example, in Ref.~\cite{no2sc}, 
it was argued that a mixture of the normal phase, made of strange 
quarks, and the two-flavor color superconducting (2SC) phase, made 
of up and down quarks, is less favorable than the color-flavor 
locked (CFL) phase after the charge neutrality condition is enforced. 

Assuming that the constituent medium-modified mass of the strange
quark is large (i.e., larger than the corresponding strange quark
chemical potential), in Ref.~\cite{SH} it was shown that neutral
two-flavor quark matter in $\beta$ equilibrium can have another
rather unusual ground state called the gapless two-flavor color
superconductor (g2SC). The appearance of this phase is directly
connected with enforcing the charge neutrality in the system. While 
the symmetry in the g2SC ground state is the same as that in the 
conventional 2SC phase, the spectrum of the fermionic quasiparticles 
is different. The order parameter of the g2SC phase is given by the
difference of the number densities of quarks participating in pairing 
(e.g., the number density of green up quarks and the number density 
of red down quarks). 

The existence of the gapless two-flavor color superconducting phase was 
confirmed in Refs.~\cite{GLW,var-appr,rusterR}, and generalized to nonzero 
temperatures in Refs.~\cite{HS,LZ}. Later it was shown, however, that a 
chromomagnetic instability develops in such a phase \cite{pi}. In view 
of this, the true ground state remains unknown. If the surface tension 
between different quark phases is sufficiently small, as suggested 
in Ref.~\cite{RR}, a mixed phase composed of the 2SC phase and the 
normal quark phase \cite{SHH} may be more favored than the gapless 
phases. If this conclusion holds after the screening effects in the 
mixed phase are properly taken into account, the mixed phase is 
likely to be the ground state.

It was also shown that a gapless CFL (gCFL) phase could appear in 
neutral strange quark matter when the strange quark mass is not 
very small \cite{gCFL,gCFL-long}. At nonzero temperature, the (g)CFL 
phase and several other new phases (e.g., the so-called dSC and 
uSC phases) were studied in Refs.~\cite{dSC,RSR,FKR}. Recently, 
however, it was claimed that the gCFL phase also has a chromomagnetic
instability \cite{pi-gCFL}.

\section{Gapless color-flavor locked phase}

At very large densities, the most favorable phase of quark matter
is the CFL phase in which up, down and strange quarks participate 
in Cooper pairing on almost equal footing \cite{cfl}. However, at 
the highest baryon densities existing in stars (which are less than 
about $10\rho_0$, where $\rho_0\approx 0.15$~fm$^{-3}$ is the nuclear 
saturation density), the CFL phase may be replaced by a less symmetric 
phase. This is because the strange quark is considerably heavier 
than the up and down quarks, and the ideal strange-nonstrange 
cross-flavor diquark pairing could be distorted. Indeed, it is 
most likely that the actual value of the strange quark mass $m_s$ 
in a dense medium is in the range between about $100\,\mbox{MeV}$ 
and $500\,\mbox{MeV}$. This is not negligible compared to the quark 
chemical potential $\mu$ which is of the order of $500\,\mbox{MeV}$ 
in the center of compact stars.

Here, we consider a Nambu-Jona-Lasinio (NJL) type model for
three-flavor quark matter with a local current-current interaction,
\begin{equation}
{\cal L} = \bar\psi\left(i\gamma^\mu\partial_\mu+\gamma^0\hat\mu
-\hat{m}\right)\psi + \frac{g^2}{2\Lambda^2}\left(\bar\psi\gamma^\mu
\frac{\lambda^A}{2}\psi\right)^2.
\end{equation}
where the color-flavor structure of the chemical potential and the mass
matrices are given by
\begin{equation}
\hat\mu = \mu + \mu_Q Q + \mu_3 T_3 + \mu_8 T_8
\end{equation}
and $\hat{m}=\mbox{diag}_{\rm flavor}(0,0,m_s)$, respectively. The 
matrices $Q$, $T_3$ and $T_8$ are the generators of mutually commuting 
electric and two color charges.

In the Cooper pairing dynamics responsible for color
superconductivity, the main effect of a non-vanishing strange quark
mass is a reduction of the strange quark Fermi momentum,
\begin{equation}
k_F^{(s)} = \sqrt{\mu^2-m_s^2} \simeq \mu-\frac{m_s^2}{2\mu}, \quad
\mbox{for} \quad m_s\ll \mu.
\label{k_F_shift}
\end{equation}
The magnitude of the reduction is approximately given by the value 
of $m_s^2/2\mu$. This quantity plays the role of a mismatch parameter
in three-flavor quark matter, which is similar to $\delta\mu\equiv
\mu_e/2$ in two-flavor quark matter \cite{SH}. This mismatch interferes 
with Cooper pairing between strange and non-strange quarks \cite{gCFL}. 

The simplest way to take into account the effect of the strange 
quark mass is to replace the chemical potential of the strange 
quark by its effective shifted value in Eq.~(\ref{k_F_shift}). This 
was the approach of Refs.~\cite{gCFL,RSR}. In this paper, as in 
Ref.~\cite{FKR}, we do not use such an approximation. The strange 
quark mass is properly taken into account.

Because of a nonzero strange quark mass, the up-down, the up-strange and 
the down-strange diquark condensates are not equal in the ground state 
\cite{gCFL,n_steiner},
\begin{equation}
\left\langle \left(\bar{\psi}^C\right)_i^a \gamma^5 \psi_j^b
\right\rangle
\sim  \phi_1 \varepsilon_{ij1} \epsilon^{ab1}
    + \phi_2 \varepsilon_{ij2} \epsilon^{ab2}
    + \phi_3 \varepsilon_{ij3} \epsilon^{ab3}
+\cdots,
\label{LL-RR-gCFL}
\end{equation}
where the ellipsis denote the terms symmetric in color and flavor. 
Although the symmetric terms are small and not crucial for the qualitative 
understanding of strange quark matter, we retain them in our analysis 
\cite{RSR}.

A nonzero value of the strange quark mass 
interferes most prominently with the pairing between the strange 
and the non-strange quarks, i.e., with the pairing described by the 
gap parameters $\phi_1$ and $\phi_2$. Because of color-flavor 
locking, preserved in the diquark condensate (\ref{LL-RR-gCFL}), this 
translates into a special role played by the blue color in the 
ground state (in QCD this is meaningful, provided a specific gauge 
fixing is done).

Starting from the massless limit ($m_s=0$) and gradually 
increasing the value
of the strange quark mass, one finds that the CFL phase stays robust until
a critical value of the control parameter $m_s^2/2\mu \simeq \Delta$
is reached \cite{gCFL}. Here, $\Delta$ is the value of the gap parameter
$\phi_1$. (Note that $\Delta\equiv \phi_1=\phi_2 \approx \phi_3$ in the 
CFL phase, see left panel in Fig.~\ref{phase-d} below.) Above the 
critical value, the charge neutrality exerts too much stress on the CFL 
phase, and a transition to a new (gapless) phase happens \cite{gCFL}.

A nice feature of the CFL phase is that it stays almost automatically
electrically neutral \cite{enforce_n}. The reason is that Cooper
pairing in the CFL phase helps to enforce equal number densities of
all three quark flavors, $n_u=n_d=n_s$. Since the sum of the charges
of up, down and strange quarks add up to zero, this insures that the
electric charge density is vanishing, $n_Q=\frac{2}{3}n_u-\frac{1}{3}
n_d-\frac{1}{3}n_s=0$. This is exactly what happens in the CFL phase
even at nonzero, but sub-critical values of the strange quark mass.

In contrast to the g2SC case, it is the color rather than the electric 
charge neutrality that plays the key role in destabilizing the CFL phase 
of three-flavor quark matter with increasing the value of $m_s^2/2\mu$.
The actual mechanism is directly related to color-flavor locking in the 
CFL ground state. Because of such a locking, the blue quarks have a 
special status in the Cooper pairing dynamics. In order to avoid the 
violation of the color neutrality by these quarks, a nonzero value of 
the color chemical potential $\mu_8\propto -m_s^2/2\mu$ is required 
\cite{no2sc}. Note that the value of $\mu_8$ is monotonically increasing
with the strange quark mass. After the stress in the quark system 
becomes too strong, the CFL phase turns into the gapless CFL phase. 
As was shown in Ref.~\cite{gCFL}, this happens when $m_s^2/2\mu \approx 
\phi_1$. In essence, the mechanism is the same as in two-flavor 
quark matter studied in Ref.~\cite{SH}.

\section{Phase diagram} 

In this section, we present the phase diagram of dense neutral 
three-flavor quark matter in the plane of temperature and $m_s^2/\mu$.
The first version of such a phase diagram was presented in Ref.~\cite{RSR}.
In Ref.~\cite{RSR}, however, the effect of the strange quark mass was 
incorporated only through a shift of the chemical potential of 
strange quarks, $\mu^{a}_{s} \to \mu^{a}_s - m_s^2/(2\mu)$ (here $a=1,2,3$ 
is the color index). Such an approach is certainly reliable at small 
values of the strange quark mass. One should check, however, whether 
the results are reliable at least qualitatively also at not very small 
values of the strange quark mass. 

The study of the phase diagram \cite{RSR} was further developed in 
Ref.~\cite{FKR} where the strange quark mass was properly
taken into account. Here, we perform a similar study using our 
original set of model parameters \cite{RSR}. As we shall see, the 
results do not differ very much from those in Ref.~\cite{RSR} even 
at rather large values of the strange quark mass. 

\begin{figure}
\begin{center}
\hbox{
\epsfxsize=2.99in
\epsfbox{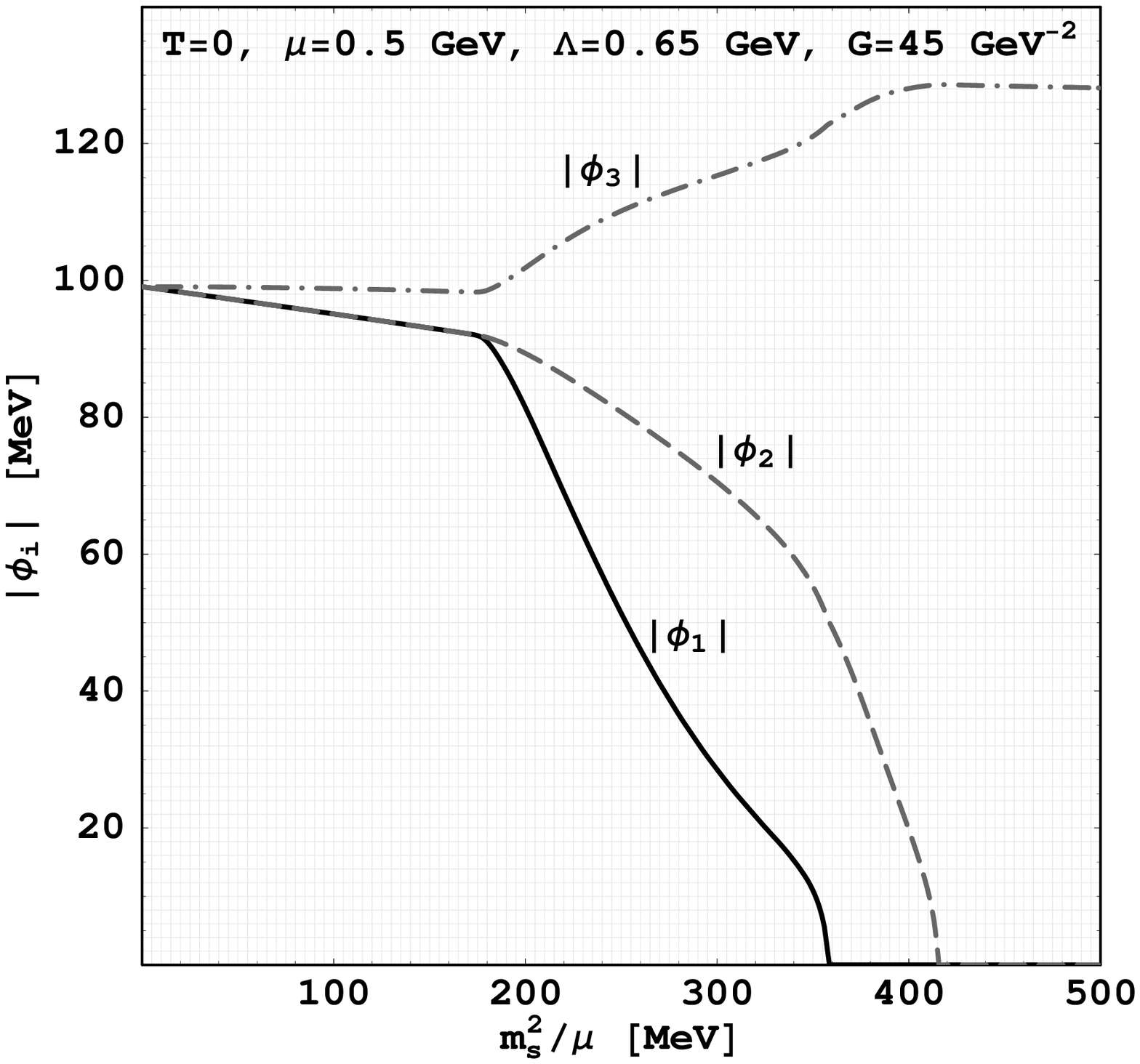}
\epsfxsize=2.99in
\epsfbox{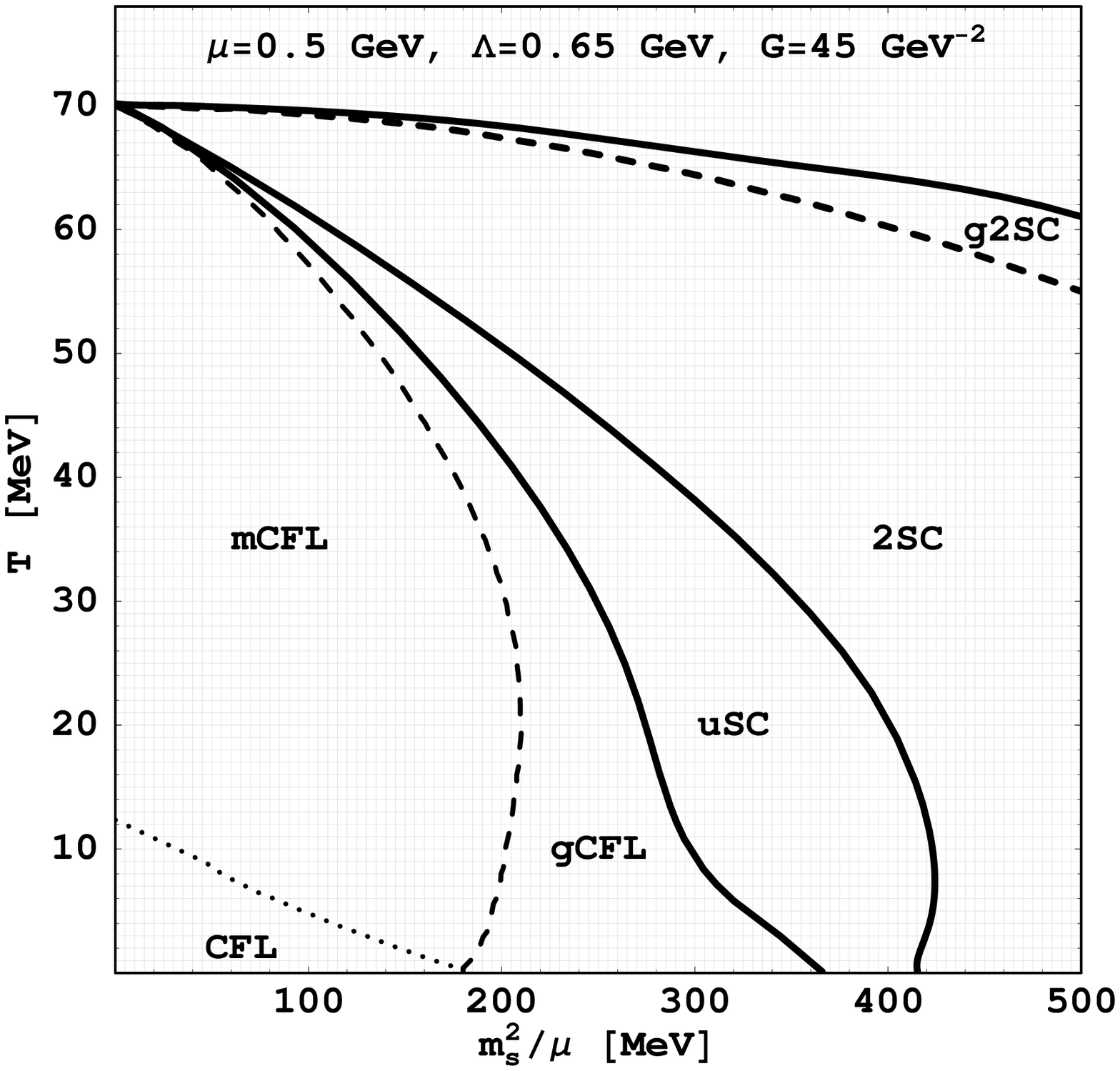}
}
\end{center}
\caption{The absolute value of the gap parameters $\phi_i$ 
     ($i=1,2,3$) as functions of $m_s^2/\mu$ at 
     zero temperature (left panel); and the phase diagram of 
     neutral three-flavor quark matter in the plane of 
     temperature and $m_s^2/\mu$  at a fixed value of the 
     quark chemical potential, $\mu=500$ MeV, (right panel).
     By definition, $G \equiv g^2/ \Lambda^2 $.}
    \label{phase-d}
\end{figure}

Let us start with the discussion of the effect of a nonzero strange
quark mass on the gap parameters. The zero-temperature results for 
the gaps as functions of $m_s^2/\mu$ are shown in the left panel of 
Fig.~\ref{phase-d}. At small strange quark mass, the ground state 
corresponds to the CFL phase. Here, the following relation between 
the three gaps holds: $\phi_1=\phi_2\approx \phi_3$ \cite{FKR}. 
At large strange quark mass, on the other hand, the three gap 
parameters are very different. As we can see from the figure, the 
qualitative change of the gaps as functions of $m_s^2/\mu$ happens at 
$m_s^2/\mu\approx 2\phi_1$. This is a consequence of the phase 
transition between the CFL and the gCFL phase \cite{gCFL}. 

As one can check, the transition point at $m_s^2/\mu\approx 2\phi_1$
corresponds to the appearance of additional gapless modes in the 
quasiparticle spectrum, justifying the name of the phase. As in the 
case of the g2SC phase, the order parameter in the gCFL phase could 
be identified with a difference of number densities of some quarks 
participating in pairing \cite{SH}. In particular, this is the 
difference of the number densities of blue down and green 
strange quarks \cite{RSR}. In Ref.~\cite{gCFL}, however, 
it was suggested to use the number density of electrons as an 
alternative order parameter. While there are no electrons in the 
CFL phase \cite{enforce_n}, there is a non-vanishing density 
of them in the gCFL phase. Thus, the corresponding transition was
called an insulator-metal transition. Note that, at nonzero 
temperatures, the corresponding transition becomes a smooth 
crossover \cite{RSR,FKR}.

The physical properties of the gCFL phase are very different from 
those of the CFL phase. The presence of gapless quasiparticle 
modes has a large effect on the thermodynamics as well as on the 
transport properties. In contrast to the CFL phase which is an 
insulator, the gCFL phase is a metal with a nonzero number 
density of electrons \cite{gCFL,gCFL-long,RSR,FKR}. Therefore, 
the electrical conductivities of the two phases are very 
different. (Note that, at low temperature, the electrical 
conductivity in the CFL phase is dominated by thermally 
excited electron-positron pairs \cite{SE}.) Also, the neutrino 
emissivity from the gCFL phase should be rather high. It is 
dominated by the $\beta$ processes involving the gapless modes. 
In contrast, the corresponding emissivity from the CFL phase is 
strongly suppressed.

Now, let us discuss how three-flavor neutral quark matter responds to 
a nonzero temperature. In general, as in Ref.~\cite{RSR}, if one starts 
from the (g)CFL phase and gradually increases the temperature, three 
consecutive phase transitions occur in the system (see right panel 
of Fig.~\ref{phase-d}):
\begin{itemize}
\item[(i)] the transition from the (g)CFL phase to the so-called uSC phase;
\item[(ii)] the transition from the uSC phase to the 2SC phase; 
\item[(iii)] the transition from the 2SC phase to the normal quark phase.
\end{itemize}
Here, the notation uSC (dSC) stands for superconducting phases in which
there are only {\em up}-down and {\em up}-strange (or up-{\em down} and 
{\em down}-strange) condensates, and there is no down-strange (or 
up-strange, respectively) condensate \cite{dSC}. 
From Eq.~(\ref{LL-RR-gCFL}) one can check that $\phi_1$ vanishes in 
the uSC phase, while $\phi_2$ vanishes in the dSC phase. 

Our results differ from those of Ref.~\cite{dSC} in that the dSC phase 
is replaced by the uSC phase in the near-critical region. This was also 
the case in our previous study \cite{RSR}. However, Ref.~\cite{FKR} 
revealed a small region of the dSC phase at temperatures close 
to the critical temperature and at small values of the strange 
quark mass. We checked that our numerical calculations produce 
qualitatively the same results when we use a set of parameters 
close to that of Ref.~\cite{FKR}. In fact, the main difference 
between the two studies is the value of the cut-off parameter in the 
NJL model. From this we conclude that the size of the dSC region in the 
phase diagram is particularly sensitive to the choice of the cut-off 
in the NJL model. 

Now, let us briefly discuss the main features of the phase diagram of 
dense neutral three-flavor quark matter in the plane of temperature 
and $m_s^2/\mu$. This is shown in the right panel of Fig.~\ref{phase-d}. 
Here, the results are plotted for a fixed value of the quark chemical 
potential, $\mu=500$ MeV. The three solid lines denote the three phase 
transitions discussed earlier. In the mean-field approximation used in 
this study, all three transitions are second order phase transitions. 
After taking into account various types of fluctuations, the nature of
some of them may change \cite{fluct}. A detailed study of this issue is,
however, outside the scope of this paper. The two dashed lines mark 
the appearance of gapless modes in the metallic CFL (mCFL) and 2SC 
phases (see Ref.~\cite{RSR} for the detailed definitions).  In 
addition, there is also an insulator-metal crossover transition 
between the CFL and mCFL phase. This is marked by the dotted line on the 
phase diagram in Fig.~\ref{phase-d}.

By comparing the results in the right panel of Fig.~\ref{phase-d}
with the corresponding phase diagram in Ref.~\cite{RSR}, we find 
that the results are qualitatively the same and even quantitatively
very similar. Thus, with our set of parameters, even a simplified 
treatment of the strange quark mass reproduces the overall 
structure of the phase diagram. It should be admitted, however, 
that this may not always be the case. For example, the simplified 
method with an effective shift of the strange quark chemical 
potential is unlikely to capture the appearance of the first order 
phase transition at small temperatures and large $m_s$ in the phase 
diagram shown in Figs.~1 and 16 of Ref.~\cite{FKR}.

\section{Conclusion}
\label{conclusions}

In this paper, we studied neutral three-flavor quark matter at large
baryon densities. We obtained the phase diagram of dense neutral 
three-flavor quark matter in the plane of temperature and $m_s^2/\mu$.
In contrast to the approximate treatment of the strange quark mass 
of Ref.~\cite{RSR}, here the mass is properly taken into account. 
The final results are very similar to those of Ref.~\cite{RSR}.

If we ignore the possibility of the chromomagnetic instability
\cite{pi,pi-gCFL} for a moment, there are two main 
possibilities for the strange quark matter ground state at $T=0$: 
the CFL and the gCFL phases in the case of small and large strange 
quark mass, respectively. The transition from the CFL to the gCFL 
phase is driven by a gradual build-up of the stress in the quark 
system due to the color neutrality condition. This stress grows 
with increasing the strange quark mass. The mechanism is directly 
related to color-flavor locking in the CFL ground state. Turning 
on the strange quark mass tends to induce an imbalance of the blue 
color in the system. This imbalance is removed by a nonzero value 
of the color chemical potential $\mu_8$ in the CFL phase. After 
reaching a critical value of the strange quark mass, $m_s^2/\mu
\approx 2\phi_1$, the CFL turns into the gCFL phase \cite{gCFL}. 
This is similar to a transition between the 2SC and the g2SC phases
which, however, is driven by the electron chemical potential,
needed to preserve electric charge neutrality in two-flavor quark 
matter \cite{SH}.

In this study we confirm the results of Ref.~\cite{RSR} regarding the 
existence of several different phases of neutral three-flavor quark 
matter at nonzero temperature. We also confirm the order in which 
they appear. In particular, we observe the appearance of the uSC 
phase as an intermediate state in melting of the (g)CFL phase into
the 2SC phase. Formally, this is different from the prediction of 
Ref.~\cite{dSC}. We find, however, that the difference is 
connected with the choice of the model parameters. In the 
NJL model with a cut-off parameter $\Lambda=800$~MeV used in 
Ref.~\cite{FKR}, there is a non-vanishing (although rather small) 
region of the dSC phase. On the other hand, in the NJL model with a 
relatively small value of the cut-off parameter (we have 
$\Lambda=653$~MeV) used in this paper, no sizeable window of 
the dSC phase is found.

Now, if one takes the chromomagnetic instability \cite{pi,pi-gCFL} 
into account seriously, there is a fundamental problem in the present
understanding of the phase diagram of neutral dense quark matter. 
Indeed, some regions of the phase diagram (see right panel of 
Fig.~\ref{phase-d}) correspond to phases that are unstable, and 
there exist no unambiguous alternatives (two of such alternatives 
were proposed in Ref.~\cite{alternatives}). Of course, it is of 
prime importance to resolve this crisis.

\section*{Acknowledgements}
The work of I.S. was supported by Gesellschaft f\"{u}r 
Schwerionenforschung (GSI) and by Bundesministerium 
f\"{u}r Bildung und Forschung (BMBF).

\end{document}